\documentstyle[prl,aps,epsf]{revtex}
\twocolumn
\tighten

\begin{document}
\draft

\twocolumn[
\hsize\textwidth\columnwidth\hsize\csname@twocolumnfalse\endcsname
\title{Superconducting Transition in Doped Antiferromagnet}
\author{Ming Shaw,$^{1,2}$ Z.Y.Weng,$^{3}$, and C. S. Ting$^1$}
\address{ $^1$
Texas Center for Superconductivity, 
University of Houston, Houston, Texas 77204\\
$^2$ National Laboratory for Superconductivity, Institute of Physics, CAS, Beijing 100080, China\\
$^3$ Center for Advanced Study, Tsinghua University, Beijing 100084, China }
\maketitle
\begin{abstract}
We investigate the superconducting transition in a doped antiferromagnet. Based on the phase string 
framework of the t-J model, an effective model describing the phase-coherence transition is obtained
and is studied through duality transformation and renormalization group treatment . We show that such a topological transition is controlled by spin excitations, 
with the transition temperature determined by a characterisitic spin excitation energy. The existence of an 
Ising-like long range order of staggered current loops is also 
discussed.
\end{abstract} 
]

\narrowtext

One of puzzles in high-$T_c$ cuprate superconductors is that the behavior of
the energy gap in the quasiparticle channel is quite different from that of
the superconducting transition temperature $T_c$. In the underdoped regime,
two scales show divergent doping dependence: the energy gap seems to
increase as the hole concentration $\delta $ is reduced, while $T_c$ itself
monotonically decreases~\cite{d99}. The former is even present above $T_c.$
By contrast, in the BCS theory of superconductivity, the quasiparticle
energy gap is simply proportional to $T_c$ at $T=0$ and vanishes above $T_c$%
. So $T_c$ seems controlled by a rather different low-energy physics in the
cuprate superconductors. For example, the well-known Uemura plot~\cite{y87}
shows the direct proportionality between $T_c$ and the phase stiffness.
Emery and Kivelson have conjectured\cite{v95}, for underdoped cuprates, $T_c$
is decided by the phase coherence~of the pairing order parameter. On the
other hand, a quantitative feature has been recently revealed by inelastic
neutron-scattering experiments. It was found\cite{d01} that $T_c$ is
correlated, in roughly a linear relation, with the characteristic spin
energy scale $E_g$ associated with a magnetic ``resonance-like'' peak ~\cite
{m98,d01}, which decreases with the doping concentration.

How to establish the connection of $T_c$ with either the phase coherence~%
\cite{v95,m198} or the spin resonance energy~\cite{weng,c99} is currently a
hot topic in various modelling, but few attempt has been made to put both
the phase coherence and spin resonance energy within a single framework in
understanding the mechanism of superconducting transition in the cuprates.
In this paper, we approach this issue by using a microscopic theory of doped
antiferromagnet based on the $t-J$ model. In such a description, one starts
from the half-filling where the antiferromagnetism is well understood. The
doping will then introduce the so-called phase string effect as doped holes
pick up sequences of nontrivial signs from the spin background during their
hopping\cite{w97}. Such a phase string effect at finite doping will destroy
the long-range antiferromagnetic spin order, leading to a sharp
``resonance-like'' peak at a doping-dependent energy scale $E_g\sim J\delta $
in the superconducting phase, characterized by the holon Bose condensation%
\cite{w98}. Due to the same phase string effect, spin excitations in the
superconducting state induce strong phase frustrations on the holon
concentration and eventually destroy the phase coherence of the latter at a
finite temperature. We will demonstrate that $T_c$ as the onset of the phase
coherence temperature is indeed proportional to $E_g$, which is not directly
associated with the ``energy gap'' in the quasiparticle channel. As a
by-product of the microscopic theory, we also show that there exists
staggered current loops in the superconducting ground state, with a hidden
broken $Z_2$ symmetry. 

We start from the t-J model. In the slave-particle representation one can
take the aforementioned singular phase string effect into account by a
decomposition\cite{w97}: $c_{i\sigma }=h_i^{\dagger }b_{i\sigma }e^{i\Theta
_{i\sigma }}$, where $h_i^{\dagger }$ and $b_{i\sigma }$ are $bosonic$ holon
and spinon operators, respectively. Using it, the local singularity of phase
strings can be ''gauged away'' from the Hamiltonian and kept in the phase
factor $e^{i\Theta _{i\sigma }}$, while the long distance topological
properties are explicitly retained in the Hamiltonian\cite{w97}. It is
believed that the resulting nontrivial topological effect of phase strings
is the key to understanding the physics of the two-dimensional (2D) t-J
model. Based on the exact reformulation of the $t-J$ model with explicitly
incorporating the phase string effect, an effective Hamiltonian was obtained~%
\cite{w98}: $H_{eff}=H_h+H_s$, in which the holon Hamiltonian is given by 
\begin{equation}
H_h=-t_h\sum\limits_{<ij>}e^{i(A_{ij}^s+\phi _{ij}^0)}h_i^{\dagger }h_j+H.c.
\end{equation}
Here the phase string effect is precisely tracked by the lattice gauge field 
$A_{ij}^s$ and $\phi _{ij}^0$, satisfying ${\sum_{\Box }}A_{ij}^s=\pi /4{%
\sum_{l\in \Box }}\sum_\sigma \sigma n_{l\sigma }^b$ and ${\sum_{\Box }}\phi
_{ij}^0=\pm \pi $ per plaquette, respectively. Note that $n_{l\sigma }^b$
denotes the spinon number operator at site $l$. So $A_{ij}^s$ will mediate
the main influence of the spinon degrees of freedom on the holon part in a
form of gauge field.

In this framework, the superconducting phase is realized by the holon Bose
condensation\cite{w98}. To study its phase transition, we shall assume that
the amplitude of the holon condensation has been formed at some
characteristic temperature $T^{*\text{ }}\geq T_{c\text{ }}$such that the
holon operator can be written as $h_i=\sqrt{\rho _h}e^{i\theta _i}$ ($\rho
_h\simeq \delta )$. Then the holon Hamiltonian can be reexpressed as 
\begin{equation}
H_h=-\rho _hg{\sum\limits_{\langle ij\rangle }}\cos [\theta _i-\theta
_j-\phi _{ij}^o-A_{ij}^s],  \label{hhc}
\end{equation}
where $g\equiv 2t_h$. In the spinon resonating-valence-bond (RVB)
background, the paired spinons will not directly contribute to the lattice
gauge field $A_{ij}^s$ according to its definition. So only thermally
excited spinons will be seen by holons through $A_{ij}^s$ in (\ref{hhc}).
Without the lattice gauge field $A_{ij}^s$, the holon Bose condensation
would occur as a conventional Kosterlitz-Thouless (KT) transition ($\phi
_{ij}^o$ as a $\pi $ flux per plaquette mainly introduces additional
frustrations to such a system). The corresponding transition temperature $%
T_{KT}$ has been estimated\cite{ln} to be over $1000K$, using the parameters
of the $t-J$ Hamiltonian, which is about one order of magnitude higher than
observed in the cuprate superconductors. In the following, we will show how $%
A_{ij}^s$ can effectively bring down the transition temperature to a value
determined by the spin characteristic energy, consistent with the experiment.

We note that the spinon part is governed by\cite{w98} $H_s=-J_s{%
\sum_{<ij>\sigma }}e^{i\sigma A_{ij}^h}b_{i\sigma }^{\dagger }b_{j-\sigma
}^{\dagger }+H.c$. At half-filling, without $e^{i\sigma A_{ij}^h}$, $H_s$
reduces to the mean-field version of Schwinger-boson representation of the
Heisenberg model ~\cite{w98}, which well captures the antiferromagnetic
correlations there. Upon doping, ${\sum_{\Box }}A_{ij}^h=1/2{\sum_{l\in \Box
}}n_l^h$ ($n_l^h$ is the holon number operator) describes fictitious
quantized $\pi $ fluxoids bound to holons and seen by spinons. So $A_{ij}^h$
will represent the doping influence on the spinon part, a consequence again
due to the phase string effect\cite{w97}. In the superconducting phase as
holons are Bose-condensed, $H_s$ has been studied in Ref. 9, where it is
shown that a resonance-like peak will emerge in the spin dynamic
susceptibility at $(\pi ,\pi )$ at a finite energy $E_g=2E_s$ with $E_s$
denoting the single spinon excitation energy. If higher energies are
neglected at low T, one has $H_s\approx E_g/2\sum_{s\sigma }\gamma _{s\sigma
}^{\dagger }\gamma _{s\sigma }$ where $\gamma _{s\sigma }$ is the Bogoliubov
operator for spinon excitations\cite{w98}. Note that $A_{ij}^h$ only depends
on the density of holons. So one expects that the peak is still present at $%
T_c<T<T^{*\text{ }}$as long as the amplitude of the holon condensation
persists. To first order approximation, in the following we shall treat $E_g$
or $E_s$ as a T-independent quantity below $T^{*}$.

Now we can write down the partition function corresponding to the present
system as follows:

\begin{eqnarray}
Z &=&{\sum\limits_{\{n_{s\sigma }^\gamma \}}}\int D\theta D^{\prime }A_\mu
^sD^{\prime }\phi _\mu ^0\times  \nonumber \\
&&e^{\beta {\sum\limits_r}\cos [\Delta _\mu \theta (r)-\phi _\mu ^o(r)-A_\mu
^s(r)]-\frac{Eg}{2T}\sum_{s\sigma }n_{s\sigma }^\gamma }  \label{z}
\end{eqnarray}
where $\beta =\rho _hg/T{\text{ , }n_{s\sigma }^\gamma \equiv \gamma
_{s\sigma }^{\dagger }\gamma }_{s\sigma }${, }${\text{and }}$the primes in $%
D^{\prime }A_\mu ^s$ and $D^{\prime }\phi _\mu ^0$ imply that $A_\mu ^s$ and 
$\phi _\mu ^0$ satisfy the constraints on ${\sum_{\Box }}A_\mu ^s$ and ${%
\sum_{\Box }}\phi _\mu ^0$ given above. For convenience, the subscript $\mu $
is introduced here to denote the link $ij$, and $\theta _i-\theta _j$ is
replaced by $\Delta _\mu \theta (r)$.

First of all, let us discuss the role of $\phi _\mu ^o(r)$. Note that the
Hamiltonian (\ref{hhc}) has the form of an extended xy model. Apart from the 
$U(1)$ symmetry, there is also an additional local $Z_2$ symmetry which
corresponds to the invariance for a transformation ${\sum_{\Box }}\phi
_{ij}^0=\pm \pi \rightarrow \mp \pi $ at each plaquette (one can realize
this by changing a link phase by $2\pi $ within each plaquette). Like the xy
model, one may reexpress the partition function (\ref{z}) in the Coulomb gas
representation through a standard duality transformation~\cite{k78}. In this
representation there are three different topological charges on the dual
lattice site, corresponding to the vorticities of a supercurrent loop
induced by the $\pi $ flux of $\phi _\mu ^o(r)$, vortices bound to spinons
through gauge field $A_{ij}^s$, and the conventional $2\pi $ vortices,
respectively. All these topological charges are coupled with each other
through long range logarithmic interactions. This lead to correlations among
the topological charges. It is clear that topological charges with opposite
vorticities have a tendency to pair at low temperature. Especially the
pairing between the opposite charges of topological vortices related to $%
\phi _\mu ^o$ means that local $Z_2$ symmetry can be broken at low
temperature~\cite{v77}. One can expect that apart from the topological
transition, there is also a Ising-like long rang order of staggered current
loop at low temperature~\cite{ff83}. Because of $\phi $$_\mu ^o$, there are
two degenerate ground states and, correspondingly, there are two low energy
modes\cite{c78}. By constructing a Landau-Ginzburg-Wilson description, one
can demonstrate that (\ref{hhc}) may be decomposed into two coupled extended
xy models. In this representation the topological transition is determined
by the extended xy terms corresponding to each low-energy mode, while the Z$%
_2$ symmetry broken is determined by the renormalized behavior of the
coupling constant $h$ between the two modes where $h$ is renormalized to
strong coupling, and the two modes are locked with relative phase $0$ or $%
\pi $~\cite{c78,m85}. Since generally the $Z_{2\text{ }}$ symmetry broken
temperature is higher than the topological transition temperature~\cite{h85}%
, in the following we mainly focus on the low-T phase without further
considering the fluctuation effects induced by $\phi _\mu ^o(r).$ The main
effect of the modes coupling due to $\phi _\mu ^o(r)$ will be represented by
the renormalization of $\beta $. One can introduce an effective $\beta
^{\prime }=\rho _hg^{\prime }/T$ with $g$ being replaced by $g^{\prime }$ to
denote the effect. A further study on the $Z_{2\text{ }}$symmetry-broken and
staggered current loop is to be given elsewhere.

The distinctive feature under this single mode approximation is the
spinon-vortex introduced by $A_\mu ^s$ with the corresponding vorticity
given by $n(r^{*})\pi \equiv \epsilon _{\nu \mu }\Delta _\nu A_\mu ^s(r)=\pi
/4{\sum_{l\in \Box }}\sum_\sigma \sigma n_{l\sigma }^b$. By means of the
standard Villain approximation\cite{v75} and duality transformation\cite{k78}%
, we can then arrive at the following form 
\begin{eqnarray}
Z &=&\sum\limits_{\{n_{s\sigma }^\gamma \}}\int D\phi e^{-\frac 1{2\beta
^{\prime }}\sum\limits_{r^{*}}[\Delta _\mu \phi (r^{*})]^2}  \nonumber \\
&&^{+2\pi i\sum\limits_{r^{*}}m(r^{*})\phi (r^{*})+\pi
i\sum\limits_{r^{*}}n(r^{*})\phi (r^{*})-\frac{E_g}{2T}\sum\limits_{s\sigma
}n_{s\sigma }^\gamma }  \label{z1}
\end{eqnarray}
where $\phi $ is the ``spin wave'' variable of the xy model, $m(r^{*})=0,\pm
1,\pm 2,...$, represent the topological charges of the ordinary $2\pi $
vortex of the xy model, and $n(r^{*})$ represents the topological charge of
the $\pi $ vortices associated with spinon-vortices, where $r^{*}$ denotes a
lattice site dual to $r$. From (\ref{z1}) we see that there are two types of
vortex perturbations for the ```spin wave'' fixed line of the system: one is
the vortex with vorticity $2\pi $ and the other is the vortex with vorticity 
$\pi $. According to Amit et. al ~\cite{a80}, the perturbation of vortex
with vorticity $n\pi $ has the scaling dimension $n^2/2$ near the critical
point. With this result we see that the $2\pi $ vortex perturbation is
irrelevant compared to the $\pi $ vortex. It will not affect the critical
behavior of the system. We can thus only focus on the effect of the $\pi $
vortices, associated with excited spinons, in the following study of
low-temperature topological transition. The last term on the right hand side
of (\ref{z1}) controls the fluctuations of $n(r^{*})$. In fact, one can
write down the relation: 
\begin{equation}
n(r^{*})\simeq {\sum_{s,\sigma }}|w_s(r^{*})|^2\sigma n_{s\sigma }^\gamma ,
\end{equation}
in which $w_s(r)$ is the single-particle wave function for spinons at $E_{s%
\text{ }}$\cite{w98} and in the coherent-state representation\cite{wen88} $%
w_s(r)\simeq c_se^{-({\bf r}-{\bf r_s})^2/4l_0^2}$, with $c_s$ the
normalized constant. In this representation the quantum numbers of the
states are denoted by the position ${\bf r}_s$ of their centers which form a
von Neumann lattice with lattice constants ${a_s}={b_s}\simeq 2l_0$ which is
a function of the density of doped hole: $l_0=\frac a{\sqrt{\pi \delta }}$.
At low temperature $T\ll E_g$, the terms with $n_{s\sigma }^\gamma =0$
dominates, and the terms with $n_{s\sigma }^\gamma =1$ can be regarded as
small corrections. Then $\sum\limits_{\{n_{s\sigma }^\gamma \}}\exp \{\pi
i\sum\limits_{r^{*}}n(r^{*})\phi (r^{*})-{E_g}/2{T}\sum\limits_{s\sigma
}n_{s\sigma }^\gamma \}\simeq \exp \{{y_b}\sum\limits_{r_s}\cos [\pi
\sum\limits_{r^{*}}\mid w_s(r^{*})\mid ^2\phi (r^{*})]\}$, where $%
y_b=1/\cosh (E_g/2T)$ ($y_b<<1$ ), in deducing it all the terms with order $%
O(y_b^n)$ ($n>1$ ) have been omitted. It is easy to understand that $y_b$
measures how easy (difficult) to excite a spinon. Finally the continuum form
of the partition function is obtained as follows (for convenience in the
following we use $r$ instead of $r^{*}$ to denote the dual lattice site):

\begin{eqnarray}
Z &=&\int D\phi e^{-\frac 12\int d^2r(\Delta _\mu \phi )^2+\frac{y_b}{%
(2l_0)^2}\int d^2r_s\cos [\frac{\pi \sqrt{\beta ^{\prime }}}{a^2}\times } 
\nonumber \\
&&^{\int d^2r\mid w_s(r)\mid ^2\phi (r)]-y\int d^2r_s[\frac{2l_0}{a^2}\int
d^2r\partial _s\mid w_s(r)\mid ^2\phi (r)]^2},  \label{z2}
\end{eqnarray}
where $a$ is the lattice constant. We have introduced a term $y\int d^2r_s[%
\frac{2l_0}{a^2}\int d^2r\partial _s\mid w_s(r)\mid ^2\phi (r)]^2$ into the
partition function, which will be generated by renormalization group (RG)
procedure discussed below with its initial value being zero. The partition
function is now parametrized by three quantities: $\beta ^{\prime }$, $y$,
and $y_b$.

Compared to the conventional xy model, the partition function (\ref{z2})
looks more complicate. The physical origin of this complexity comes from the
spinons, which are at centers of the $\pi $ vortices. Note that an excited
spinon does a cyclotron motion\cite{w98}, and in our coherent-state
representation its distribution function $\mid w_s(r)\mid ^2$ has an
attenuation radius $l_0$. So the $\pi $ spinon-vortex has a finite
vortex-core of radius $l_0$, which is reflected in (\ref{z2}) through $\mid
w_s(r)\mid ^2$. We shall treat the problem by means of the Wilson's RG
~analysis\cite{s94,k79}. In this RG procedure we first divide $\phi $ into
the high energy part and low energy part, respectively: $\phi _\Lambda =\phi
_{\Lambda ^{\prime }}+h$, where $\phi _\Lambda (r)=\int_{0<p<\Lambda
}d^2p/(2\pi )^2\phi (r)\exp [i{\bf p}.{\bf x}]$, $\phi _{\Lambda ^{\prime
}}(r)=\int_{0<p<\Lambda ^{\prime }}d^2p/(2\pi )^2\phi (r)\exp [i{\bf p}.{\bf %
x}]$, $\Lambda ^{\prime }=\Lambda -d\Lambda $, and $\Lambda =1/a$ is the
momentum cut off; then the high energy part $h$ is averaged out by means of
the cumulant expansion; finally we rescale the system so as to restore the
original cutoff. Because of the presence of a characteristic length scale $%
2l_0$ of the vortex core, the RG analysis is separated into two steps; we
first treat local physics within the length scale $2l_0$, then study the low
energy and long wave-length physics beyond such a scale. After some algebra,
following recursion relations can be obtained 
\begin{eqnarray}
dl_0 &=&-l_0\frac{da}a,  \label{zz1} \\
dy_b &=&-\frac{\beta ^{\prime }}4\frac 1{(\Lambda l_0)^2}y_b\frac{da}a,
\label{zz2} \\
dy &=&c_1{\beta ^{\prime }}^2y_b^2\frac 1{(\Lambda l_0)^2}\frac{da}a,
\label{zz3}
\end{eqnarray}
where $c_1=0.018\pi ^2$. From (\ref{zz1})-(\ref{zz3}), the effective
parameters at the length scale $2l_0$ can be determine: $y_b^{\prime
}=y_b(\Lambda _f)=\frac{e^{\beta ^{\prime }(\pi \delta /4-1)/2}}{cosh(E_s/T)}
$, $y^{\prime }=y(\Lambda _f)=2c_1\beta ^{\prime }\frac{1-e^{\beta ^{\prime
}(\pi \delta /4-1)}}{cosh^2(E_s/T)}$, where $\Lambda _f=1/(2l_0)$. With the
length scale $2l_0$ being reduced to our ``new'' lattice constant $a$, the
effective attenuation radius of $\mid w_s(r)\mid ^2$ also ``shrinks'' from $%
l_0$ to $a/2$. So the effective range of $\mid w_s(r)\mid ^2$ is within the
unit plaquette of the new lattice now, which can be reasonably treated as
the $\delta (r)$ function, such that $\frac 1{a^2}\int d^2r\mid w_s(r)\mid
^2\phi (r)\simeq \phi (r_s)$. The kinetic term in (\ref{z2}) can be then
rewritten as $-\frac 12(1+2y^{\prime })\int d^2r[\partial \phi (r)]^2$.
After the variable change $\phi \rightarrow \frac 1{\sqrt{1+2y^{\prime }}}%
\phi $, we finally obtain 
\begin{equation}
Z=\int D\phi e^{-\frac 12\int d^2r(\Delta _\mu \phi )^2+\frac{y_b^{\prime }}{%
a^2}\int d^2r\cos [\pi \sqrt{\beta ^{\prime \prime }}\phi (r)]},  \label{z3}
\end{equation}
where $\beta ^{\prime \prime }=\beta ^{\prime }/(1+2y^{\prime })$. Equation (%
\ref{z3}) is exactly the partition function of sine-Gordon model\cite
{a80,k79,j78} with effective parameters $y_b^{\prime }$ and $\beta ^{\prime
\prime }$.

\begin{figure}[h]
\hspace{1cm}
\centerline{\epsfxsize=4.1in \epsfbox{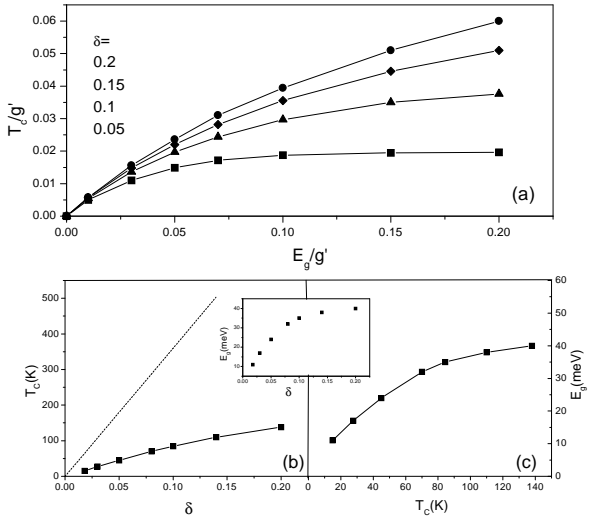}}
\caption[]{(a) $T_c$ as a function of the spin resonance-like energy $E_g$ at
different dopings, from top to bottom corresponding to $\delta=$0.2,
0.15, 0.1, and 0.5 respectively. 
(b) $T_c$ versus $\delta$ (solid) with using $E_g(\delta)$
shown in the inset and $g^{\prime}=0.2$$eV$. The dashed line represents the
ordinary KT transition temperature without considering spinon-vortices. (c) $%
E_g(\delta)$ versus $T_c(\delta)$ is ploted for the same parameters as in
(b). The Inset: $E_g(\delta)$ versus $\delta$ obtained in Ref.
$\protect\cite{w99}$. }
\end{figure}

This model has been well studied, and the transition temperature
(i.e., $T_c$), at which pairing between spinon-vortices and -antivortices
dissolves, is determined by the following equation\cite{k79}:

\begin{equation}
\frac \pi 4\beta ^{\prime \prime }-2=y_b^{\prime }/2.  \label{z4}
\end{equation}

Numerical results are shown in Fig. 1. $T_c$ as a function of $E_g$ is
plotted at several doping concentrations in Fig. 1(a). It shows that $T_c$
increases monotonically with $E_g$ and saturates at larger $E_g$. According
to Fig. 1(a), $T_c/g^{\prime }$ linearly scales with $E_g/g^{\prime }$ at
small $E_g/g^{\prime }$ where $T_c/E_g$ does not depend on $g^{\prime }$,
with the ratio ($\sim 1/4-1/5)$ being only weakly doping dependent. So $T_c$
essentially is determined by the characteristic spin resonance-like energy $%
E_g$. For instance, plugging in the experimental value of $E_g=41$ $meV$ for
the optimal doping YBCO compound\cite{m98}, one estimates $T_c\sim 100$ $K$,
very close to the experimental value. Since $E_g$ as a function of $\delta $
has already been obtained in the same framework\cite{w99} as re-plotted in
the inset of Figs. 1(b) and (c) (with $J_s=0.1$ $eV$ in Ref.\cite{w99}), one
can use such a calculated $E_g(\delta )$ to determine $T_c$ vs. $\delta $ as
shown in Fig. 1(b) (solid curve). For comparison, the dashed line in Fig.
1(b) represents $T_{KT}$ $(\simeq \pi \delta g^{\prime }/2)$ without
including the spinon-vortices due to $A_{ij}^s.$ Here we choose $g^{\prime
}\sim 2t_h=0.2$ $eV,$ and find $T_{KT}\simeq 510$ $K$ at $\delta =0.14$
while $T_c=107K.$ As noted before, $T_c$ is not sensitive to $g^{\prime },$
but $T_{KT}$ does. So if we take $g^{\prime }=0.5$ $eV$ at the same $\delta
=0.14$, $T_c$ increases to $164K$ but $T_{KT}$ reaches $1276$ $K$.
Therefore, $E_g$ effectively brings $T_c$ down to the right order of
magnitude as the consequence that the spinon-vortices instead of the
conventional $2\pi $ vortices control the superconducting phase coherence
transition. In Fig. 1(c), $E_g(\delta )-$ $T_c(\delta )$ is shown, which
also are both qualitatively and quantitatively in good agreement with the
experimental results\cite{d01}.

In conclusion, we have established a quantitative theory of superconducting
transition based on an effective spin-charge separation description of the
doped AF Mott insulator. The underlying physics is that the phase coherence 
transition is controlled by {\em thermal spin excitations},
which substantially reduce $T_c$ from $T_{KT}$
to a fraction of $E_g/k_{B.}$ It resolves the long-standing issue of how $%
T_c $ can be quantitatively connected to the characteristic spin energy
scale in a doped AF Mott insulator. The obtained $T_c-\delta $ and $E_g-T_c$
relations are in good agreement with those observed in the cuprate
superconductors, lending a strong support for the experimental relevance of
the spin-charge separation theory.

{\bf Acknowledgments} -We acknowledge useful discussions with Dr. Y. Chen.
This work is partially supported by the State of Texas through ARP Grant No.
3652707, the Texas Center for Superconductivity at University of Houston,
and the Robert A. Welch Foundation. M.S. is also partially supported by the
Ministry of Science and Technology of China (NKBRSF-G19990646), 
and NSFC project 10074078.


\begin{references}
\bibitem{d99}  H.Ding, et al., Nature {\bf 382},51(1996).

\bibitem{y87}  Y.J.Uemura et al., Phys. Rev. Lett. {\bf 58}, 2691(1987).

\bibitem{v95}  V.J.Emery and S.Kevelson, Nature {\bf 374},434(1995).

\bibitem{d01}  P.Dai et al., Phys. Rev. B{\bf 63}, 054525(2001).

\bibitem{m98}  H.F.Fong et al., Phys. Rev. Lett. {\bf 78}, 713(1997);\newline
P.Bourges, in {\it {The gap symmetry and fluctuations in high temperature
superconductors}}, edited by J.Bok et al.(Plenum Press, 1998), see also
cond-mat/9901333.

\bibitem{m198}  M.Franz, et al., Phys.Rev.B.{\bf 58}, 14572(1998).

\bibitem{weng}  Z.Y. Weng, in {\em High Temperature Superconductivity},
edited by S.E. Barnes, {\it et al}. (AIP, Woodbury, New York 1999).

\bibitem{c99}  J.P.Carbotte, et al., Nature{\bf 401}, 354(1999).

\bibitem{w97}  Z.Y.Weng et al, Phys. Rev. B{\bf 55}, 3894(1997).

\bibitem{w98}  Z.Y.Weng, D.N.Sheng, and C.S.Ting, Phys. Rev. Lett. {\bf 80},
5401(1998); Phys. Rev. B{\bf 59}, 8943(1999).

\bibitem{ln}  P.A. Lee and Nagaosa, Phys. Rev. B{\bf 46}, 5621 (1992).

\bibitem{k78}  L.P.Kadanoff, J. Phys. A{\bf 11},1399(1978).

\bibitem{v77}  J.Villain, J. Phys. C{\bf 10},4793,(1977);\newline
S.Teitel et al., Phys. Rev. B{\bf 27}, 598(1983); G.R.Santiago et al., Phys.
Rev. Lett. {\bf 68}, 1224(1992); P. Olsson, Phys. Rev. Lett.{\bf 75},
2758(1995).

\bibitem{ff83}  The only difference of our model from fully frustrated XY
(FFXY) model~\cite{v77} is the presence of the gauge field $A_{ij}^s$. The
role of $A_{ij}^s$ is to introduce $\pi $ votices associated with spinons.
We note that the FFXY model without $A_{ij}^s$ has been also studied
recently by Z. Nussinov in cond-mat/0107339.

\bibitem{c78}  M.Y.Choi et al., Phys. Rev. B{\bf 31}, 4516(1985).

\bibitem{m85}  M.Yosefin et al., Phys. Rev. B{\bf 32},1778(1985).

\bibitem{h85}  T.C.Halsey., J. Phys. C{\bf 18}, 2437(1985).

\bibitem{v75}  J.Villain, J. Phys. C{\bf 36}, 581(1975).

\bibitem{a80}  D.Amit et al., J. Phys. A{\bf 13}, 585(1980).

\bibitem{wen88}  X.G.Wen and A.Zee, Phys. Rev. B{\bf 41}, 240(1990);
E.I.Rashba et al., Phys. Rev. B{\bf 55}, 5306(1997).

\bibitem{s94}  R.Shankar, Rev. Mod. Phys. {\bf 66}, 129(1994).

\bibitem{k79}  J.B.Kogut, Rev. Mod. Phys. {\bf 51}, 659(1979).

\bibitem{j78}  J.V.Jose et al., Phys. Rev. B{\bf 16}, 1217(1977).

\bibitem{w99}  Z.Y.Weng, D.N.Sheng, and C.S.Ting, Phys.Rev.B{\bf 61},
12328(2000).
\end{references}
\end{document}